\documentclass[showpacs, pra,twocolumn,preprintnumbers ,amsmath, amssymb, superscriptaddress, aps]{revtex4-2}
\usepackage{color}
\usepackage{amsmath,amssymb}
\usepackage{pifont}
\usepackage{amssymb}  
\usepackage{bbold}
\usepackage{float}
\usepackage{subfloat}
\usepackage[caption=false]{subfig}
\usepackage{tikz}
\usepackage{makecell}
\usepackage{subfig}
\usepackage{pifont}   
\usepackage{graphicx} 
\graphicspath{{Figures/}}
\usepackage{dcolumn}  
\usepackage{bm}       
\usepackage{multirow} 
\usepackage{placeins}
\usepackage[colorlinks]{hyperref}
\usepackage{mathtools}

\newcommand{\ket}[1]{\left|{#1}\right\rangle}
\newcommand{\bra}[1]{\left\langle{#1}\right|}

\newcommand{\eqfitpage}[1]{\resizebox{\linewidth}{!}{$#1$}}

\begin{document}
\title{Floquet Hofstadter butterfly in trilayer graphene with a twisted top layer  
}
\date{\today}
\author{Nadia Benlakhouy}
\email{benlakhouy.n@ucd.ac.ma}
\affiliation{Laboratory of Theoretical Physics, Faculty of Sciences, Choua\"ib Doukkali University, PO Box 20, 24000 El Jadida, Morocco}
\author{Ahmed Jellal}
\email{a.jellal@ucd.ac.ma}
\affiliation{Laboratory of Theoretical Physics, Faculty of Sciences, Choua\"ib Doukkali University, PO Box 20, 24000 El Jadida, Morocco}
\affiliation{Canadian Quantum  Research Center,
204-3002 32 Ave Vernon,  BC V1T 2L7,  Canada}
	\author{Hocine Bahlouli}
	\affiliation{Physics Department, King Fahd University
		of Petroleum $\&$ Minerals,
		Dhahran 31261, Saudi Arabia}
	\begin{abstract}
		
		The magnetic field generated Hofstadter butterfly in  single-twist trilayer graphene (TLG) is investigated using circularly polarized light (CPL) and longitudinal light emanating from a waveguide. We show that  single-twist TLG has two distinct chiral limits in the equilibrium state, and the central branch of the butterfly splits into two precisely degenerate components. The Hofstadter butterfly appears to be more discernible. We also discovered that CPL causes a large gap opening at the central branch of the Hofstadter butterfly energy spectrum and between the Landau levels, with a clear asymmetry corresponding to energy $E = 0$. We point out that for right-handed CPL, the central band shifts downward, in stark contrast to left-handed CPL, where the central band shifts upward. Finally, we investigated the effect of longitudinally polarized light, which originates from a waveguide. Interestingly, we observed that the chiral symmetries of the Hofstadter butterfly energy spectrum are broken for small driving strengths and get restored at large ones, contrary to what was observed in twisted bilayer graphene.
		 
\end{abstract}
	\maketitle
\section{Introduction}
\begin{figure}[ht]
	\centering
\fbox{\subfloat[]{\includegraphics[width=0.46\linewidth]{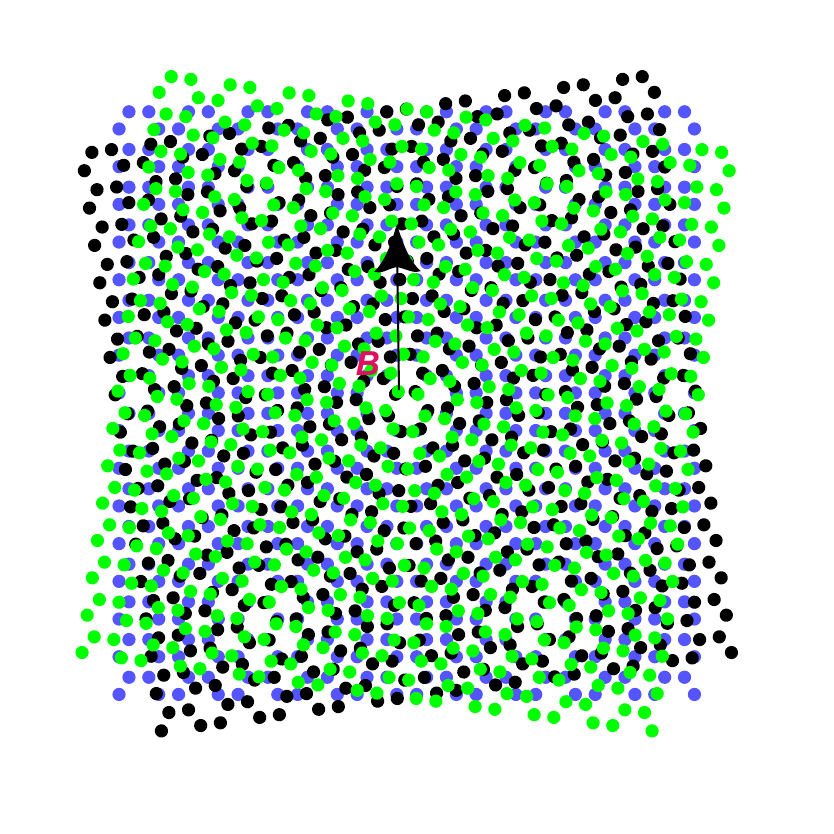} \label{fig1:TTLG_combined_figa}}
\subfloat[]{\includegraphics[width=0.49\linewidth]{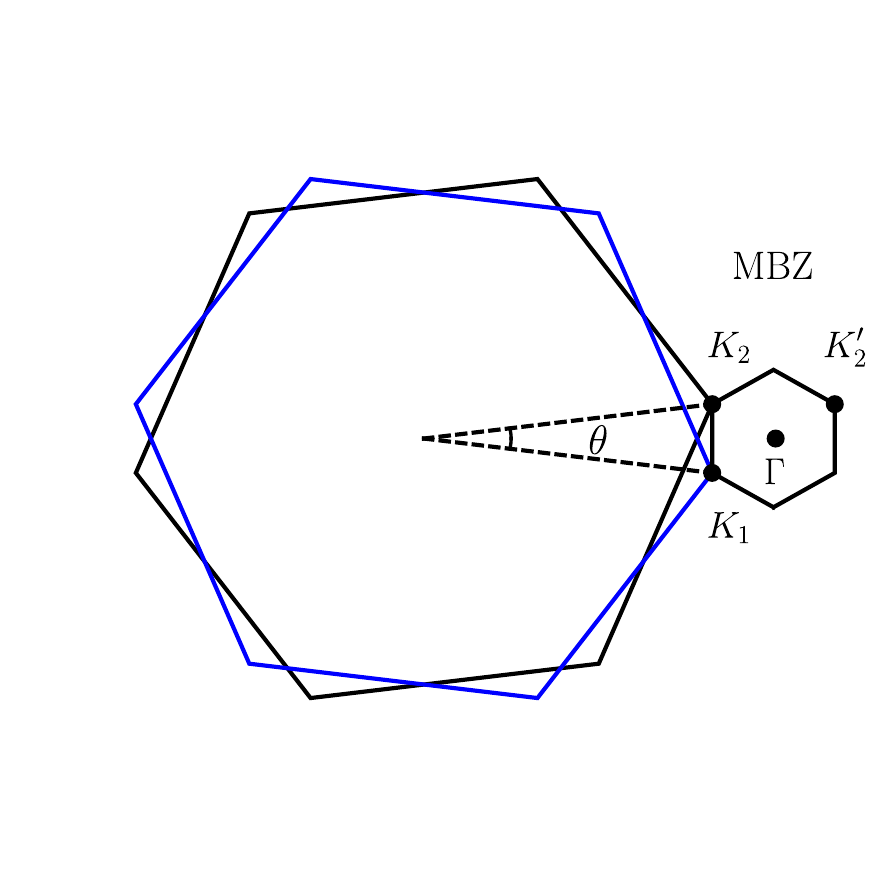}\label{fig1:TTLG_combined_figb}}}
\caption{(a) Illustration of the moir\'{e} pattern of
 twisted trilayer graphene (TTLG). (b) A schematic illustration of the moir\'{e} Brillouin zone (MBZ).}
\end{figure}

Following the discovery of superconductivity and correlation-induced insulators, physicists have taken a keen interest in investigating further physical properties of twisted multi-layer graphene \cite{macdonald2019bilayer, andrei2020graphene, kennes2021moire, balents2020superconductivity}. Most of the observed exotic phases were caused by a small twist angle between successive graphene layers, more specifically at a magic angle, where an isolated flat energy band appeared \cite{ bistritzer2011moire, dos2012continuum}. Twisted bilayer graphene (TBLG)    
\cite{cao2018unconventional, cao2020tunable,navarro2022first,naumis2021reduction,shen2020correlated, liu2020tunable, chebrolu2019flat,koshino2019band,lee2019theory,haddadi2020moire,culchac2020flat,zhang2019nearly, wong2020cascade, lu2019superconductors, PhysRevLett.128.217701, PhysRevLett.129.076401, herzog2022magnetic, yu2022correlated}, twisted trilayer graphene (TTLG) \cite{park2021tunable,hao2021electric,lopez2020electrical, assi2021floquet, li2019electronic} a variety of  other multi-layer arrangements \cite{kerelsky2021moireless,rubio2020universal} were among the most thoroughly studied structures, both theoretically and experimentally. The small twist between two or three layers of graphene produces a long-wavelength moir\'{e} pattern as described in Refs \cite{liu2014evolution, cao2016superlattice, kim2017tunable,li2010observation,cao2018unconventional} see Fig. \ref{fig1:TTLG_combined_figa} in particular. During the last few years, these moir\'{e} pattern systems have attracted considerable interest because they constitute a test bed for the occurrence of a variety of anomalous and highly correlated phases.
The recent studies of TTLG have largely focused on the electronic structure \cite{li2019electronic}, the importance of topological flat energy bands near the magic angles \cite{po2018origin, zou2018band, yuan2018model, kang2018symmetry, koshino2018maximally, hejazi2019multiple, po2019faithful}, and the rich optical properties \cite{zuo2018scanning, correa2014optical,morell2013electronic, qiao2014plane, mora2019flatbands,aoki2007dependence, bao2017stacking, kuzmenko2009determination, wu2018theory, partoens2006graphene}. 
In this context, the moir\'{e} and Floquet techniques have recently been integrated to obtain theoretical predictions regarding distinct topological phases in moir\'{e} materials \cite{li2010observation, assi2021floquet, topp2019topological, vogl2020floquet, lu2021valley, rodriguez2020floquet,   hejazi2019landau}. Floquet's theory has been exploited in studies concerned with the electromagnetic interaction of external light sources with materials where light intensity and polarization have been used to control the transport properties in these devices that are based on twisted multi-layer graphene systems.

The objective of this research is to find out more about the magnetic field generated Hofstadter butterfly in single-twist TLG subjected to various types of light, including circularly polarized light (CPL) and longitudinal light emanating from a waveguide. The famous Hofstadter butterfly shows up when a two-dimensional periodic electron system is subjected to a perpendicular magnetic field, which then exhibits fractal patterns in their energy spectrum \cite{hofstadter1976energy}. The influence of light on the Hofstadter butterfly energy spectrum has recently attracted the physics community's curiosity  \cite{wang2015atomic,wang2009quantum,lawton2009spectral,wang2013kicked,lababidi2014counter,zhou2014floquet,ding2018quantum,wackerl2019driven,kooi2018genesis, zhao2022floquet}. Previous research concentrated on the kicked Harper model as well as laser-irradiated monolayer graphene (MLG) in the presence of a uniform perpendicular magnetic field $B$  \cite{wang2015atomic,wang2009quantum,lawton2009spectral,wang2013kicked,lababidi2014counter}. In our recent work in TBLG \cite{benlakhouy2022chiral}, we found that in the non-equilibrium situation, in the presence of light, the magnetic field generated Hofstadter butterfly exhibits a much richer structure. Recent efforts focused on investigating single-twist BLG and single-twist  TLG under the effects of two kinds of light polarization, CPL and longitudinally polarized light originating from a waveguide \cite{vogl2020effective, vogl2020floquet, assi2021floquet}. CPL case was treated using a rotating frame (RF) Hamiltonian. This latter was designed to treat both strong and weak perturbations in the high to intermediate frequency ranges \cite{vogl2020effective}. In addition, this Hamiltonian is valid in the ranges where the ordinary Van Vleck approximation fails. As a result, the computations of the quasi-energy levels were greatly improved.
In single-twist TLG, the design of the crystalline stacking configuration, as well as the selection of which of the three layers is to be twisted, generates a distinct physical system with different characteristics \cite{aoki2007dependence, bao2017stacking}.  We focus here on the top single-layer twist with $\text{ABA}$ stacking. We will look at the chiral case,  such as TBLG, which is controlled by the interlayer coupling parameters  $\omega_0=0$ for $\text{AA}$-bilayer graphene and $\omega_1\neq 0$ for AB-bilayer graphene. 
It was discovered that the last hopping term plays a critical role in the creation of the flat energy bands at the magic angle \cite{tarnopolsky2019origin}.
The main difference here is that we find that $\omega_0$-type hopping is a key factor in the creation of the Hofstadter butterfly in TBLG \cite{benlakhouy2022chiral}. This kept us motivated to check this feature in single-twist TLG. We start by first studying angles larger than the magic angle and intermediate twist angle, then restricting our attention to the magic angle. Furthermore, we are expecting to obtain a rich light-induced structure of the Hofstadter butterfly in  single-twist TLG.

The remainder of this work is structured in the following manner. We describe single-twist TLG  in Sec. \ref{MODEL HAMILTONIANS}, and we base our discussion on the main features of the equilibrium state, beginning with the equilibrium model. In particular, we investigate which of the single-twist TLG's  hopping processes has the greatest impact on the Hofstadter butterfly. In addition, we demonstrate how the various hopping terms affect symmetry in relation to the central branch or $E=0$, and we investigate the impact of the twist angle on this feature. In Sec. \ref{Circulary polarized light}, a non-equilibrium case, we first discuss the effect of circularly polarised light. The second type of light polarization, longitudinal light emanating from a waveguide, is discussed in Sec. \ref{Waveguide light}. Last, we summarize our main findings and present our conclusions.

\section{Equilibrium Studies}
\label{MODEL HAMILTONIANS}
\subsection{Theoretical approach}
To begin with, we adopt the continuum model Hamiltonian for  single-twist TLG  in Refs. \cite{li2019electronic, assi2021floquet} with twisted top layer (TTL). The TTL can be captured by twisting the middle-bottom and top layers in opposite directions.  The effective model of single-twist TLG  projected onto the $+K$ valley reads as follows:
\small\begin{equation}
\mathcal{H}=\begin{pmatrix}
	-iv_{\text{F}}(\boldsymbol{\sigma}_{-\theta/2}\cdot\boldsymbol{\nabla}) & T_{12}(\boldsymbol{r}) & 0 \\
		T_{12}^{\dagger}(\boldsymbol{r}) & 	-iv_{\text{F}}(\boldsymbol{\sigma}_{\theta/2}\cdot\boldsymbol{\nabla}) & T_{23} \\
		0 & T_{23}^{\dagger} & -iv_{\text{F}}(\boldsymbol{\sigma}_{\theta/2}\cdot\boldsymbol{\nabla})
	\end{pmatrix} \label{TTLG-Ham},
\end{equation}
where the MLG tight-binding Hamiltonians reflecting the intralayer hopping of the $l$-th layer are depicted by the diagonal blocks in Eq. (\ref{TTLG-Ham}) 
\begin{equation}
h_{\ell}(\theta_{\ell})=-iv_{\text{F}}(\boldsymbol{\sigma}_{\theta_\ell}\cdot\boldsymbol{\nabla}),
\end{equation}
where the Fermi velocity in each graphene layer is represented by
$v_{\text{F}}=10^{6} \mathrm{~m} / \mathrm{s}$. 
$\boldsymbol{\sigma}=(\sigma_x,\sigma_y)$
represent Pauli matrices acting on sublattice. TTL can be specified using parameters like $\theta_1=-\theta/2$, and $\theta_2=\theta_3=\theta/2$. The rotated Pauli matrices are given by
\begin{equation}
	\boldsymbol{\sigma}_{\theta_\ell}=\left(\cos(\theta_\ell)\sigma_x-\sin(\theta_\ell)\sigma_y,
	\cos(\theta_\ell)\sigma_x+\sin(\theta_\ell)\sigma_y\right).
\end{equation}
The off-diagonal block matrix elements in (\ref{TTLG-Ham}) represent interlayer hopping. They are given by
\begin{align}
&	T_{12}(\boldsymbol{r})=\sum_{j=1}^{3} e^{-i \boldsymbol{q}_{j} \cdot \boldsymbol{r}} T_{j},
\\
&T_{23}=\sum_{j=1}^{3} T_{j},
\end{align}
where $\boldsymbol{q}_{1}=k_\theta(0, -1)$,   $\boldsymbol{q}_{2,3}=k_\theta(\pm\sqrt{3},1)$ are  the  MBZ's  nearest neighbor vectors with the modulation $k_{\theta}=2k_D\sin(\frac{\theta}{2})$. The Dirac momentum is represented by $k D=4\pi/3a_0$, with $a_0=2.46  \textup{~\AA}$ the lattice constant.  In the case of TTL, $T_{23}$ is position independent \cite{li2019electronic,assi2021floquet}. The matrices $T_j$ take into account tunneling between sublattice and depend on the type of stacking.  \cite{li2019electronic}. They have the following structure
	\begin{equation}
	\begin{gathered}
		T_{j}^{\mathrm{AB}}=[T_{j}^{\mathrm{BA}}{ }]^{\dagger}=\begin{pmatrix}
			\omega_0	e^{i \frac{2 j \pi}{3}} &\omega_1 \\
			\omega_1	e^{-i \frac{2 j \pi}{3}} & \omega_0e^{i \frac{2 j \pi}{3}}
	\end{pmatrix},
	\end{gathered}\label{Eq: interlayer-hoppinf-matrices}
\end{equation} 
	\begin{equation}
		\begin{gathered}
			T_{j}^{\mathrm{AA}}=\begin{pmatrix}
				\omega_0 & \omega_1e^{-i \frac{2 j \pi}{3}} \\
				\omega_1e^{i \frac{2 j \pi}{3}} & \omega_0
		\end{pmatrix}.
			\end{gathered}
	\end{equation}
Because AB and BA stacking are more energetically favorable than AA stacked regions \cite{vogl2020floquet, li2020floquet, katz2020optically}, we added a parameter $\omega_i$ to the tunneling term to represent the effect of relaxation. 
Additionally, $\text{AA}$ and $\text{AB}$ regions offer distinct interlayer-lattice constants \cite{guinea2019continuum}.  
In our case, we take these variables into account in the hopping amplitudes by considering $(\omega_0,~\omega_1)=(80~\text{meV}/80~\text{meV},~110~\text{meV})$ for $(\text{AB/BA}, \text{AA})$-type hopping amplitude.
In a relaxed lattice, distortions can be detected in this way \cite{assi2021floquet, vogl2020floquet, katz2020optically} if the next neighboring layer interactions are neglected, making them comparable to those in TBLG.

\subsection{Interlayer hopping and the Hofstadter butterfly}
\label{sec:analysis_eq_param} 
\begin{figure*}
	\centering
	\subfloat[ ]{\includegraphics[width=0.47\linewidth]{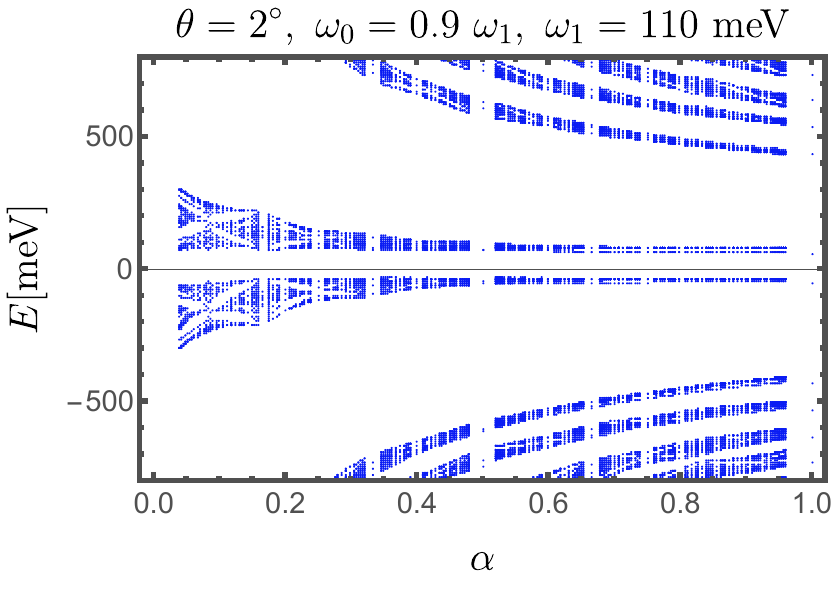}\label{fig:TTLG_combined_figa}}
	\subfloat[ ]{\includegraphics[width=0.47\linewidth]{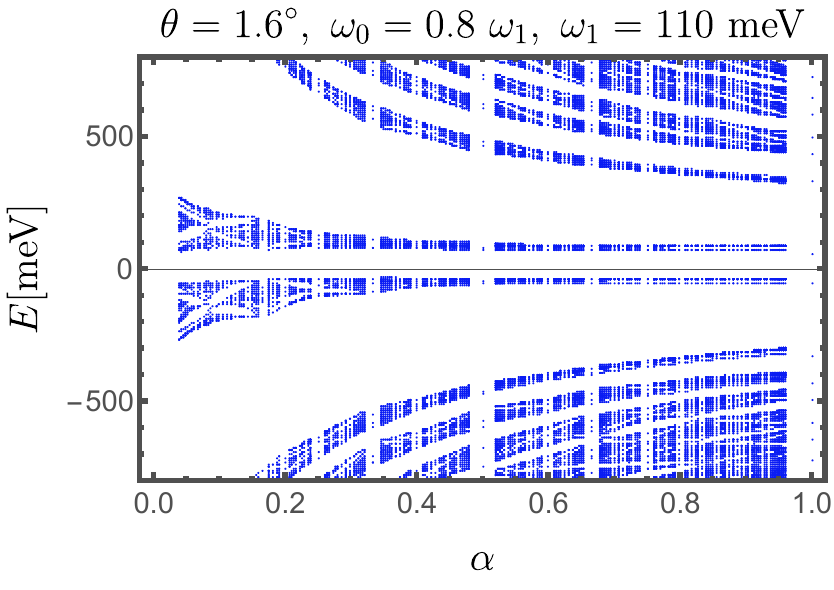}	\label{fig:TTLG_combined_figb}}\\
		\subfloat[ ]{\includegraphics[width=0.47\linewidth]{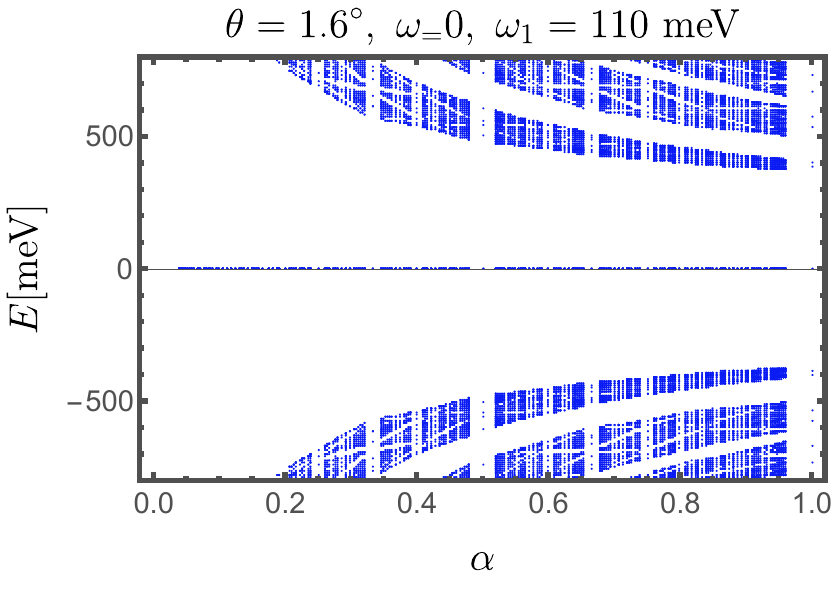}\label{fig:Hofstadter_BF__Func_InterlayerHoppingc}}
	\subfloat[ ]{\includegraphics[width=0.47\linewidth]{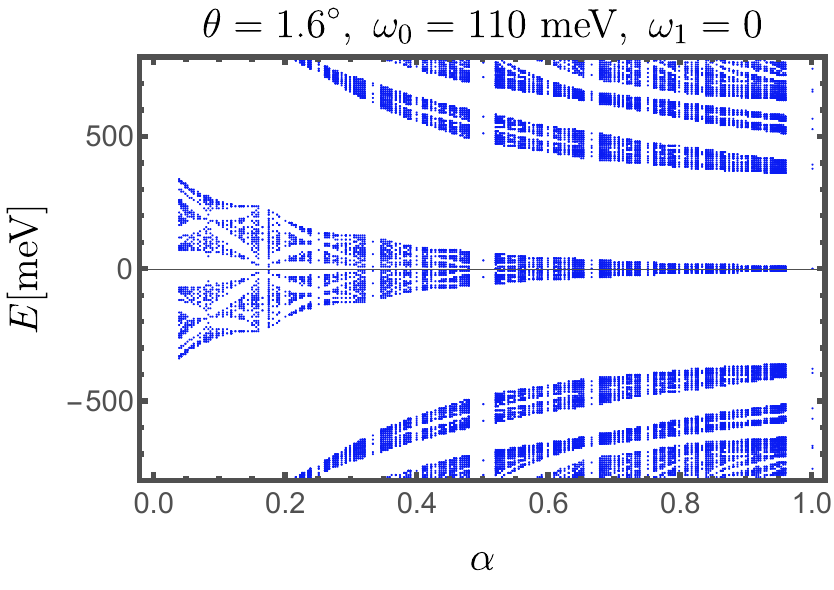}	\label{fig:Hofstadter_BF__Func_InterlayerHoppingd}}
\caption{(Color online) The numerically generated Hofstadter butterfly spectrum for single-twist TLG with parameters (a) ($\theta,\omega_0,\omega_1$)=($2^{\circ}, 0.9\omega_1, 110\ \text{meV}$), (b) ($\theta,\omega_0,\omega_1$)=($1.6^{\circ}, 0.8\omega_1, 110 \ \text{meV}$), (c) ($\theta,\omega_0,\omega_1$)=($1.6^{\circ}, 0, 110\ \text{meV}$), (d) ($\theta,\omega_0,\omega_1$)=($1.6^{\circ}, 110 \ \text{meV}, 0$).} \label{fig:TTLG_combined_fig}
\end{figure*}

The energy spectrum was calculated numerically using the assumption that \cite{bistritzer2011moire}  
\begin{equation}
N_{\mathrm{max}} \approx 2\left[\frac{\max \left(a_{0}\gamma_{\text{RF}}, \omega_1\right)} {\omega_{c}}\right]^{2}.
\end{equation}
The center of our plot, which corresponds to $E=0$, or more precisely, $n_{LL} = 0$, splits into two precisely degenerate components and shifts to higher and lower energies $69.075~\text{meV}\precsim E[\text{meV}]\precsim-62.3541~\text{meV}$ at an intermediate twist angle of $\theta=2^{\circ}$. Moreover, for small enough magnetic fields, mini gaps as wide as $332.141~\text{meV}$  started to appear between LLs.  According to $T_ {12} $, Eq. \ref{eq:ThopmagbasT_12} (see App. \ref{appendix A}), every single Landau level (LL) is divided into $q$ sub-bands and linked together. To better understand the Hofstadter butterfly in TTLG  $\text{ABA}$ TTL devices, we propose a second twist angle, which we prefer to call the magic angle $\theta_M=1.6^{\circ}$ with $\omega_0/\omega_1\simeq 0.8$. Here the Hofstadter butterfly structure appears, see Fig. \ref{fig:TTLG_combined_figb}, and the band gaps are minimized for the small magic angles. Previous studies on the square, honeycomb, and kagome lattices \cite{du2018floquet} and TBLG \cite{benlakhouy2022chiral} investigated a rich symmetry with $\alpha = 0.5$ and a reflection symmetry with $E = 0$. In contrast to the  single-twist TLG, we notice that the reflection symmetry concerning the axis of energy $E = 0$ is somewhere missing due to the interlayer hopping terms.

The next step in our equilibrium investigation will be to determine whether $\omega_0$ or $\omega_1$ interlayer hopping processes are more important than the other for the appearance of our butterfly in  single-twist TLG, as done in TBLG studies \cite{benlakhouy2022chiral}. In Fig. \ref{fig:Hofstadter_BF__Func_InterlayerHoppingc}, we show the Hofstadter butterfly in the chiral model \cite{tarnopolsky2019origin}, which is equivalent to $\omega_0=0$. As expected, our plot's center merges into a zero-energy line. This is because graphene has a single sublattice with the lowest LL, and the parameters in $T(\mathbf{r})$ that connect sublattices are directly proportional to $w_1$.
Consequently, the lowest LL is unaffected by these terms. In comparison to higher levels, this is a significant difference, which persists in both sublattices.
As previously stated, we take $V=T(\mathbf{r})$ as a result of perturbation, then the lowest LLs eigen-bi-spinors for $T(\mathbf{r})=0$ satisfy $\left|L_{01}\right\rangle=(0,|0\rangle, 0,0)$ or $\left|L_{02}\right\rangle=(0,0,0,|0\rangle)$. The first energy correction as a consequence of this can be expressed as $\left\langle L_{02}|V| L_{02}\right\rangle=0$ for $w_{0}=0$. As a result, the lowest LL remains unsplit. It's worth noting that the other LLS also start to collapse on each other more than in the case of TBLG \cite{benlakhouy2022chiral}.  For example, we can observe that $ \omega_0 $ influences the strength of the lowest LL splitting. The $\omega_0$-type hoppings then become less important in  single-twist TLG \cite{carr2019exact} for small magic angles. 
Now consider the opposite situation, $\omega_1=0$ and $\omega_0=110$ meV, as shown in Fig. \ref{fig:Hofstadter_BF__Func_InterlayerHoppingd}. Remarkably, we recognize that in these circumstances, it is this term that leads the center LL to split and the Hofstadter butterfly to appear. Tarnopolsky and colleagues \cite{tarnopolsky2019origin} investigated  the lower relevant term that allows the flat band to appear.

\section{Circularly polarized light}
\label{Circulary polarized light}
In this part, we will look at how the circularly polarized light (CPL) affects the Hofstadter butterfly in single-twist TLG.
\subsection{Effective Hamiltonian}
We assume CPL is transmitted perpendicular to the single-twist TLG  at frequency $\Omega$ and driving strength $ A $. Eventually, light comes into play through the usual minimal substitution \cite{vogl2020effective, assi2021floquet, dehghani2015out}
\begin{align}
 &k_{x} \rightarrow k_{x}(t)=k_{x}-A \cos (\Omega t),\\
 & k_{y} \rightarrow \tilde{k}_{y}= k_{y}-A \sin (\Omega r).
\end{align}
As a result, we obtain a time-periodic Hamiltonian $H(\mathbf{x}, \mathbf{k},t) = H(\mathbf{x}, \mathbf{k},t+2\pi/\Omega)$. It is well known that such a Hamiltonian can be accurately replaced by an effective time-independent Hamiltonian \cite{vogl2020effective}. 
A numerically advantageous and less costly approach would be helpful in the current situation. Therefore, let us quickly examine how to derive an effective time-dependent description and what new physical phenomena can emerge from it. Converting a regularly driven Hamiltonian into an RF is a non-perturbative way of finding the effective time-independent Hamiltonian. The following transformation can be used for this purpose \cite{vogl2020effective}: 
\begin{equation}
H_R =U^{\dagger}(t)\left(H-i \partial_{t}\right) U(t).
\end{equation}
A Hamiltonian is generated by a subsequent time average if a suitable frame is selected. This is a more honest description than Hamiltonians generated by van Vleck or Floquet-Magnus approximations, which are common high frequency regimes \cite{rodriguez2021low}. It was revealed that using a well-chosen unitary transformation \cite{vogl2020effective, assi2021floquet} that for single-twist TLG subjected to CPL, a highly accurate effective Hamiltonian will result
\begin{widetext}
\begin{equation}
	\mathcal{H}(\boldsymbol{r},t)=\begin{pmatrix}
	v_{\mathrm{RF}} R(\theta_{1}) \mathbf{k}\cdot \boldsymbol{\sigma}-\Delta_{\mathrm{RF}} \sigma_{3}& \tilde{T}_{12}(\boldsymbol{r}) & 0 \\
		\tilde{T}_{12}^{\dagger}(\boldsymbol{r}) &		v_{\mathrm{RF}} R(\theta_{2}) \mathbf{k}\cdot \boldsymbol{\sigma}-\Delta_{\mathrm{RF}} \sigma_{3} & \tilde{T}_{23} \\
		0 & \tilde{T}_{23}^{\dagger} & 		v_{\mathrm{RF}} R(\theta_{3}) \mathbf{k}\cdot \boldsymbol{\sigma}-\Delta_{\mathrm{RF}} \sigma_{3}
\end{pmatrix}\label{eq:TBG_RotFrame},
\end{equation}
\end{widetext}
where $R(\theta)$ denotes the rotation matrix. The Fermi velocity has been affected and is now equal to
\begin{equation}
	v_{\mathrm{RF}}=v_F J_{0}\left(-\frac{6 \gamma}{\Omega} J_{1}\left(\frac{2 A a_{0}}{3}\right)\right) J_{0}\left(\frac{2 A a_{0}}{3}\right),
\end{equation}
$J_0$ is the zeroth Bessel function of the first kind. 
Light also provokes the system to create a band gap, which is expressed as
\begin{equation}
	\Delta_{\mathrm{RF}}=-\frac{3 \gamma}{\sqrt{2}} J_{1}\left(\frac{2 A a_{0}}{3}\right) J_{1}\left(-\frac{6 \sqrt{2} \gamma}{\Omega} J_{1}\left(\frac{2 A a_{0}}{3}\right)\right).
\end{equation}
Interlayer tunneling matrices Eq. \ref{Eq: interlayer-hoppinf-matrices} are also modified. Now if we express $T_{j}$ as $T_{j}=\sum_i T_{j,i}\sigma_i$, where $T_{j,n}$ are expansion coefficients, we get the new hopping matrices $\tilde{T}_{j}$ 
\begin{equation}
	\tilde{T}_{j}^{\mathrm{\text{AB}}}=[\tilde{T}_{j}^{\mathrm{BA}}{ }]^{\dagger}=\sum_i T_{j,i}\tilde \sigma_i ,
\end{equation}
with the matrices
\begin{align}
		&\tilde{\sigma}_{1,2}=J_{0}(\nu) \sigma_{1,2},\\
		&\tilde{\sigma}_{0,3}=\sigma_{0,3}+\left(J_{0}(\sqrt{2} \nu)-1\right)\left[\sigma_{0,3} \sin ^{2}\left(\frac{\theta}{2}\right)\pm\frac{i}{2} \sigma_{3} \sin \left(\theta\right)\right].
	\end{align}
where $ \nu=(-6\gamma /\Omega) J_{1}\left(2 A a_{0} / 3\right)$, and $\sigma_{1,2,3}$ are the Pauli matrices, and $\sigma_0$ is the  identity matrix. 

\subsection{Numerical Results}
\begin{figure*}
	\centering
	\subfloat[]{\includegraphics[width=0.47\linewidth]{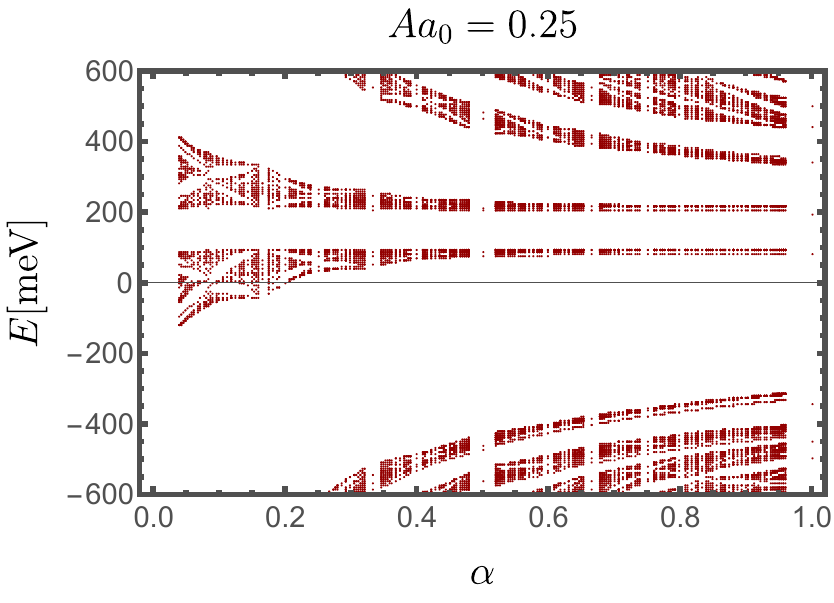}\label{fig:circ_pol_light_butterflya}}	\subfloat[ ]{\includegraphics[width=0.47\linewidth]{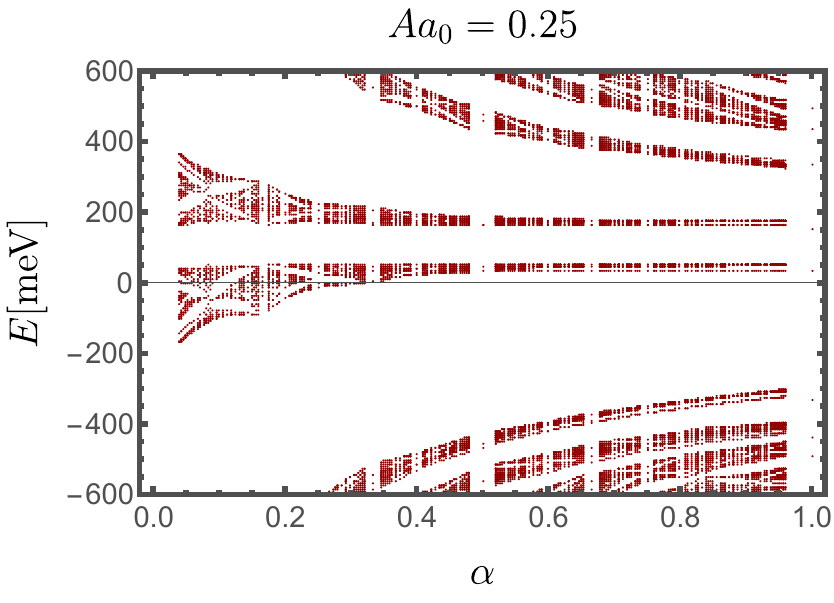}\label{fig:circ_pol_light_butterflyb}} \\	\subfloat[ ]{\includegraphics[width=0.47\linewidth]{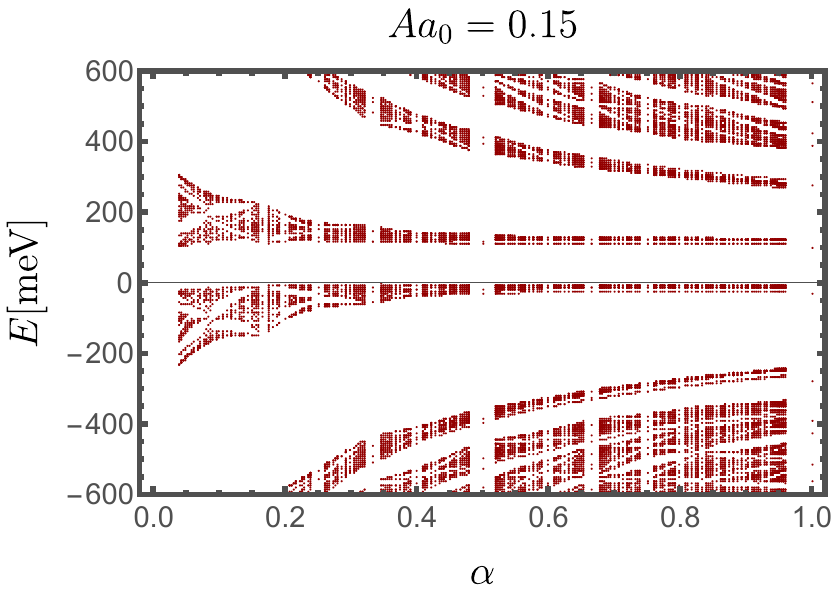}\label{fig:circ_pol_light_butterflyc}}
	\subfloat[ ]{\includegraphics[width=0.47\linewidth]{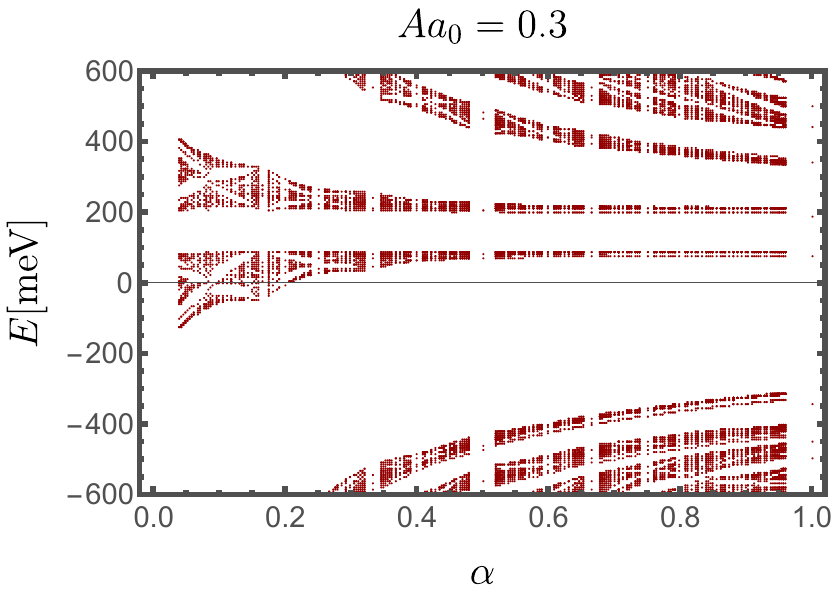}\label{fig:circ_pol_light_butterflyd}}\\
	\subfloat[ ]{\includegraphics[width=0.47\linewidth]{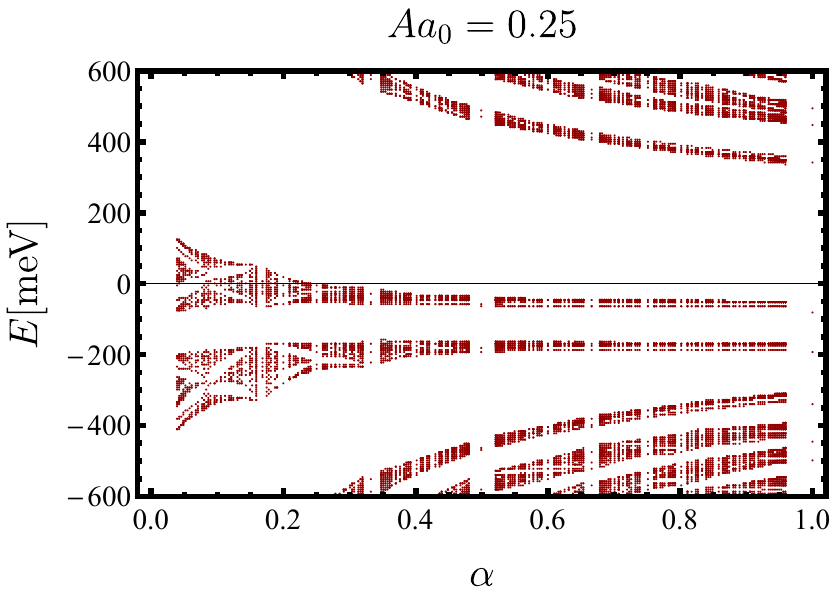}\label{fig:circ_pol_light_butterflye}}
	\subfloat[ ]{\includegraphics[width=0.45\linewidth]{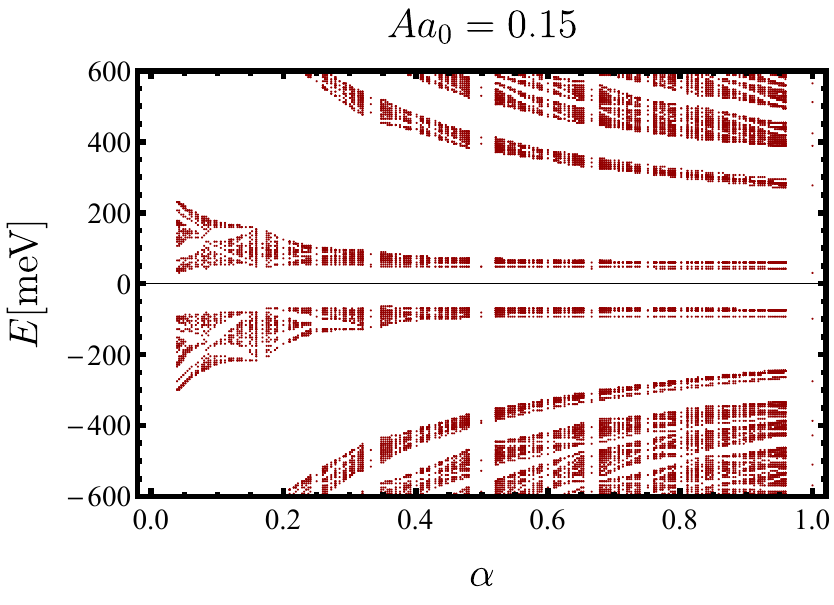} \label{fig:circ_pol_light_butterflyf}}
	\caption{(Color online) The numerically generated Floquet Hofstadter butterfly spectrum subjected to a right-handed CPL with 
		parameters:  $\omega_0=0.8~\omega_1$, $\omega_1=110~ \text{meV}$, (a) ($\theta, Aa_0, \Omega$)=($1.8^{\circ},0.25,2\gamma$), (b) ($\theta, Aa_0, \Omega$)=($1.8^{\circ},0.25,3\gamma$), (c) ($\theta, Aa_0, \Omega$)=($1.6^{\circ},0.15,3\gamma$), and  (d) ($\theta, Aa_0, \Omega$)=($1.8^{\circ},0.3,3\gamma$). Energy spectrum of the Floquet Hofstadter butterfly subjected to left-handed CPL with the parameters:  (e) ($\theta, Aa_0, \Omega$)=($1.8^{\circ},0.25,2\gamma$), and (f) ($\theta, Aa_0, \Omega$)=($1.6^{\circ},0.15,3\gamma$).}	\label{fig:circ_pol_light_butterfly}	
\end{figure*}
\begin{figure*}
	\centering
	\includegraphics[width=0.47\linewidth]{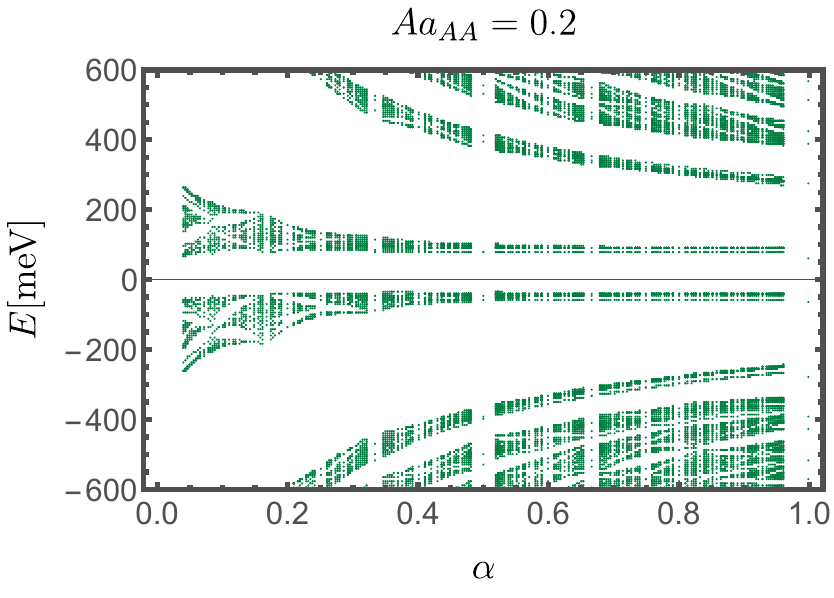}
	\includegraphics[width=0.47\linewidth]{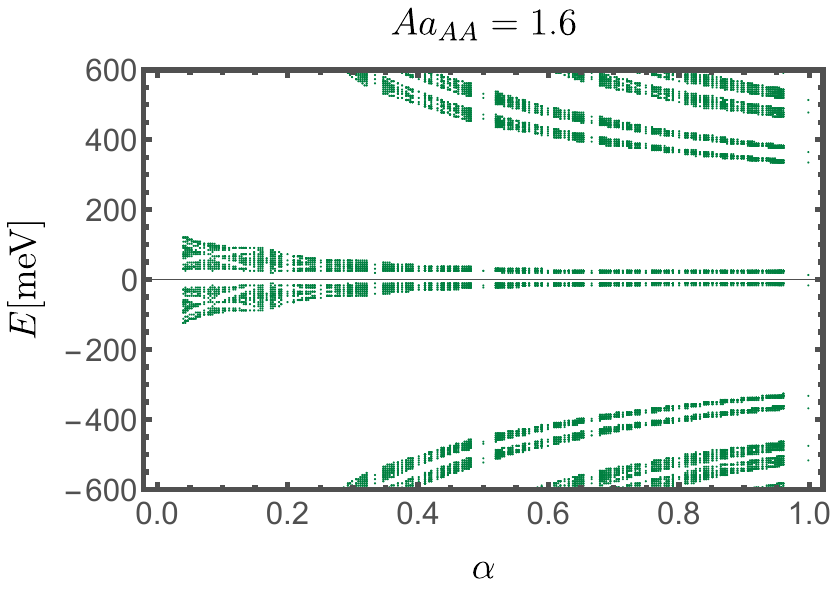}
	\includegraphics[width=0.47\linewidth]{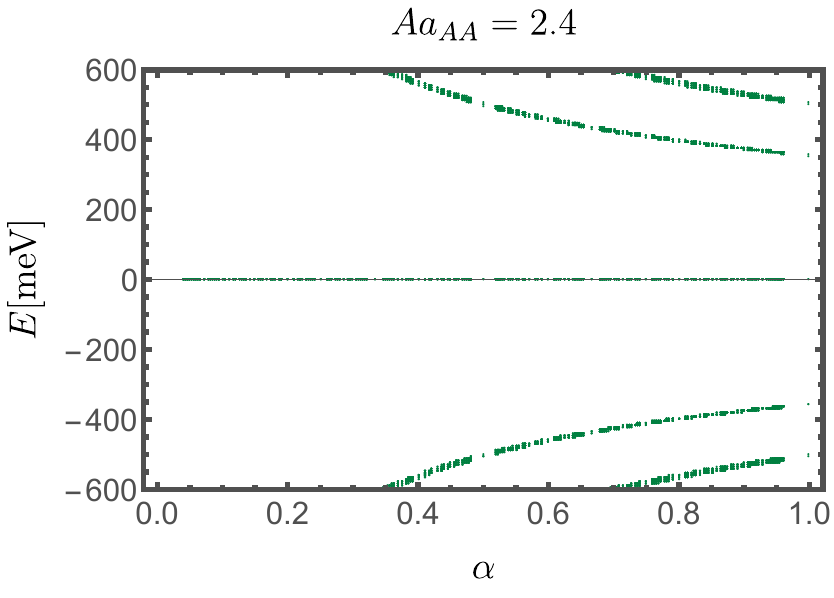}
	\includegraphics[width=0.47\linewidth]{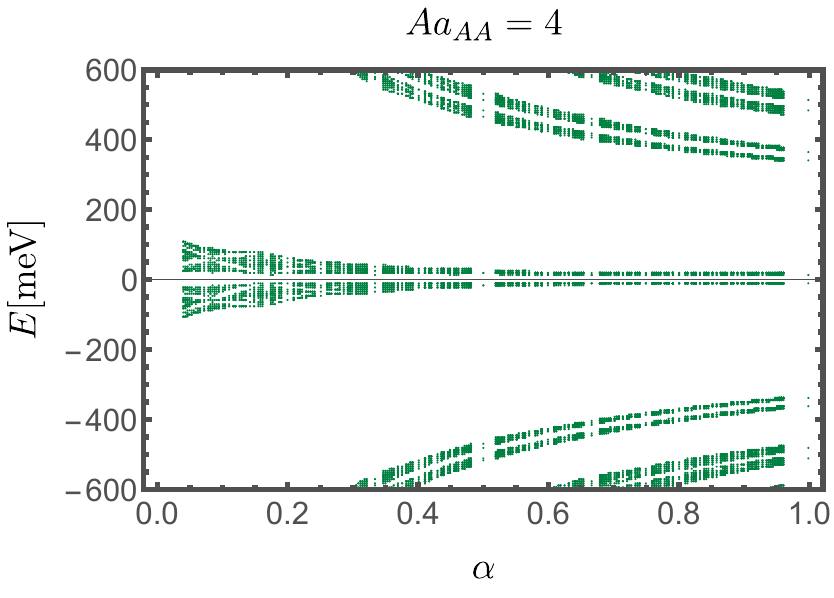}
	\label{fig:hofstadter_waveguide_light}
	\caption{(Color online) The numerically generated Floquet Hofstadter butterfly spectrum subjected to a waveguide light. The following  driving strengths have been used $Aa_{\text{AA}} = 0.2$ to $Aa_{\text{AA}} =4$ in addition to the parameters $\gamma = 2364$ meV, and $\theta=1.6^{\circ}$.}\label{fig:hofstadter_waveguide_light}
\end{figure*}
In this section, we look at how CPL influences the Hofstadter butterfly in single-twist TLG. To achieve this, in Fig. \ref{fig:circ_pol_light_butterfly}, we plotted the Hofstadter butterfly resulting from this form of light. For certain twist angles ($\theta=1.6^{\circ}/\theta= 1.8^{\circ}$), driving strength $Aa_0=0.15$, $Aa_0=0.25$, $Aa_0=0.3$, and driving frequency $\Omega=2\gamma$, and $\Omega=3\gamma $, CPL has an interesting effects. Indeed, the associated energy levels split as we increase the driving strength and the driving frequency. The $0$-th LLs of the butterfly's central branch move upwards to higher energies and are separated as shown in Fig. \ref{fig:circ_pol_light_butterflya}.
As a result, the spectrum appears to be asymmetrical with respect to $E=0$ because the band-gap $\Delta_{\text{RF}}$ violates  chiral symmetry. Compared to TBLG \cite{benlakhouy2022chiral} we note that the gaps between LLs increase in single-twist TLG. As a result, it's an appealing choice in strongly correlated phases, since interactions are expected to dominate in this case, as discussed in \cite{assi2021floquet}. We highlight that the central LL is shifted upward in Figs. Figs. \ref{fig:circ_pol_light_butterflya}, \ref{fig:circ_pol_light_butterflyb}, \ref{fig:circ_pol_light_butterflyc}, and \ref{fig:circ_pol_light_butterflyd} since we are considering right-handed CPL. Additionally, switching from a right-handed to a left-handed CPL results in the substitution  
\begin{equation}
	\Delta_{\text{RF}}\longmapsto -\Delta_{\text{RF}}.
\end{equation}
We plot the Hofstadter butterfly subject to left-handed CPL in
 Fig. \ref{fig:circ_pol_light_butterflya} and \ref{fig:circ_pol_light_butterflyb} with the same values as in  Fig. \ref{fig:circ_pol_light_butterflya} and \ref{fig:circ_pol_light_butterflyb}. We observe that the central Landau level has been pushed downward.

\section{Waveguide light}
\label{Waveguide light}
Following our study of CPL in the previous section, we will look at the impact of longitudinal light originating from a waveguide on our TTLG spectrum.

\subsection{Theoretical Approach}

We will now consider longitudinal light emanating from a waveguide,  as a second type of light. In this situation, the waveguide's boundary conditions allow light to have longitudinal components $\mathbf{A}=\operatorname{Re}\left(e^{i k_{z} z-i \Omega t}\right) \hat{z}$ to occur in a vacuum  (More details on the derivation can be found in \cite{vogl2020floquet} or the most popular references on electromagnetism \cite{jackson1999classical}).The effect of this type of light can be investigated at the tight-binding level through a Peirls substitution 
\begin{equation}
	t_{i j} \rightarrow t_{i j} e^{ \left(-\int_{\mathbf{r}_{i}}^{r_{j}} \mathbf{A}\cdot d\mathbf{l} \right)},	
\end{equation}
where $\mathbf{A}$ is the vector potential. In the continuum, Hamiltonian hopping terms refer to $\omega_i$, which is now
\begin{equation}
w_{i} \rightarrow w_{i} e^{ \left(-\int_{r_{i}}^{r_{j}} \mathbf{A}\cdot d\mathbf{l}\right)}.
\end{equation}	
This phenomenon can be incorporated in the high frequency domain of our continuum model by modifying interlayer couplings as shown below
\begin{equation}
	\begin{aligned}
		&\omega_{1} \rightarrow \tilde{\omega}_{1}=J_{0}\left(\left|Aa_{A B} \right|\right) \omega_{1}, \\
		&\omega_{0} \rightarrow \tilde{\omega}_{0}=J_{0}\left(\left|Aa_{A A} \right|\right) \omega_{0},
	\end{aligned}
	\label{eq:wireplacements}
\end{equation}
\begin{figure}[ht]
	\centering
	\includegraphics[width=0.90\linewidth]{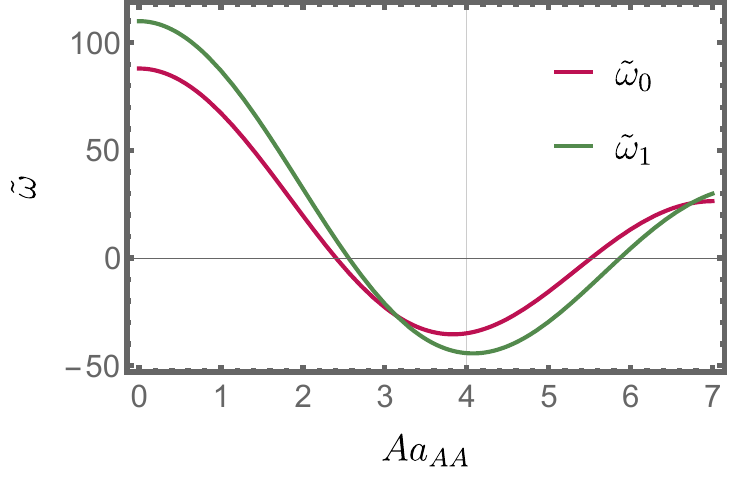}
	\caption{(Color online) Plot of the renormalized hopping amplitude $\tilde{\omega}_0$ and $\tilde{\omega}_1$ as function of the deriving strength $Aa_{AA}$.}
	\label{fig:renormalized_hopping_amplitude}
\end{figure}
\noindent where $a_{\text{AA}} = 0.36$ nm and $a_{\text{AB}} = 0.34$ nm are interlayer distances in the regions $\text{AA}$, and $\text{AB}$ respectively. 
\subsection{Numerical Results}
Next, we will start examining how the waveguide light affects the Hofstadter butterfly in single-twist TLG using our numerical results. 
Similarly to TBLG \cite{benlakhouy2022chiral}, we consider a range of distinct values $Aa_{\text{AA}}$ for our unit-less driving strength. It should be remembered that
\begin{equation}
	\frac{Aa_{\text{AB}}}{Aa_{\text{AA}}}=\frac{a_{\text{AB}}}{a_{\text{AA}}}.
\end{equation}
Fig. \ref{fig:hofstadter_waveguide_light} considers various values ranging from $Aa_{\text{\text{AA}}}=0.2$ to $4$. We notice that as $Aa_\text{\text{AA}}$ increases, the individual LL splitting decreases at first, and then increases. 
To better understand these splittings, we plot the renormalized hopping amplitudes $\tilde{\omega}_0$ and $\tilde{\omega}_1$ as a function of the deriving strength $Aa_{\text{AA}}$ in 
Fig.~\ref{fig:renormalized_hopping_amplitude}. 
As shown, the hopping amplitudes $\tilde{\omega}_0$ and $\tilde{\omega}_1$ initially decrease to be minimal for small values of the derived strength $Aa_{\text{AA}}$. Both quantities increase after $Aa_{\text{AA}}=4$, but $\tilde{\omega}_{0}$ becomes more important than $\tilde{\omega}_1$. In our case, this is modulated by the Bessel function $J_0$, which determines the magnitude of level splitting. 
 We conclude that Bessel functions influence the interlayer hoppings in single-twist TLG as it was also noticed in \cite{benlakhouy2022chiral}. 
Of course, this finding allows us to go one step further and conclude that the two chiral models $\omega_1=0$ or $\omega_0=0$ can both be generated using this type of light. Particularly, if $Aa_{\text{AA}}=j_{0,n}$ is the $n$-th zero Bessel function $J_0(x)$, $\omega_0$ is completely set to $0$. Moreover, $ \omega_1 $ is $0$ unless $Aa_{\text{AA}}=(a_{\text{AA}}/a_{\text{AB}})j_{0,n}$. We notice that the gap opening in this situation starts to close at $Aa{\text{A}}=1.6$. Under these circumstances, and in contrast to the case of CPL, with respect to $E=0$, mirror symmetry of our energy spectrum remains intact.

\section{Conclusion}
\label{conclusion}

The current study investigated the magnetic field induced Floquet Hofstadter butterfly spectrum in a top twisted ABA stacked trilayer graphene  in the presence of a perpendicular magnetic field. We first investigated the characteristics of the equilibrium state of the system in the absence of light, and after that, we went on to study different non-equilibrium situations. We specifically considered the presence of circularly polarized light and waveguide linearly polarized light. In the equilibrium state, the butterfly's central branch splits into two precisely degenerate components, and for small twist angles, such as $\theta_M=1.6^{\circ}$, our butterfly becomes more discernible.

Afterward, we have identified that the effect of interlayer coupling $\omega_0$ in the $\text{AA}$ stacking type hopping terms is much more important than the interlayer $\omega_1$ in the $\text{AB/BA}$ stacking type on the appearance of the Hofstadter butterfly. We also came to the conclusion that single-twist TLG has two separate chiral limits, similarly to TBLG. In the non-equilibrium case, in the presence of a CPL, we showed how this type of light causes large gap openings at the central branch of the Hofstadter butterfly with clearly discernible asymmetry with regard to energy $E = 0$. In addition, for right-handed CPL, the central band shifts downward, in stark contrast to left-handed CPL, where the central band shifts upward. We also investigated the effect of longitudinally polarized light emanating from a waveguide on the Hofstadter butterfly spectrum in single-twist TLG. We showed that in the case of waveguide light, chiral symmetries are broken at small driving strengths and restored at large driving strengths, contrary to previous observations in TBLG.\\

{\bf Acknowledgment}:
The authors deeply appreciate discussions with Michael Vogl on the subject matter of this paper.

\bibliographystyle{unsrt}
\bibliography{mybib}
\appendix
\section{CALCULATION OF HOFSTADTER BUTTERFLY}\label{appendix A}
In the presence of a perpendicular magnetic field, $\boldsymbol{B}=B \hat{\boldsymbol{z}}$, we substitute $\hat{\boldsymbol{p}} \rightarrow \hat{\boldsymbol{p}}+e \boldsymbol{A}$. We use the Landau gauge $\boldsymbol{A}=B(-y, 0)$ and rewrite the intralayer Hamiltonian as
\begin{equation}
	h(\theta / 2)=\omega_{c}\left[\frac{\sigma_x+i\sigma_y}{2}e^{i\theta/2}a+\frac{\sigma_x-i\sigma_y}{2} e^{-i\theta/2}a^{\dagger}\right]
\end{equation}
where $\omega_{c}=\sqrt{2} v_{\text{F}} / \ell_{b}$ is the cyclotron energy, $ \ell_{b}=\sqrt{1 / e B}$ is the magnetic length, and
\begin{equation}
	a^{\dagger}=\frac{\ell_b}{\sqrt{2}}\left[p_{x}-e B y+i p_{y}\right],
\end{equation}
\begin{equation}	
	a=\frac{\ell_b}{\sqrt{2}}\left[p_{x}-e B y-i p_{y}\right]
\end{equation}
are the raising and lowering operators of LL index. 
It is straightforward to demonstrate that  $[a, a^{\dagger}] = 1$, and they act on
 the Landau-level $n$-th eigenstate as follows
\begin{align}
&	a\ket{n}=\sqrt{n}\ket{n},
\\
&	a^{\dagger}\ket{n}=\sqrt{n+1}\ket{n+1}.
\end{align}
The Hamiltonian can be diagonalized numerically using the LLs basis  $\ket{ L,n,\alpha, y_c}$ where $L=1, 2, 3$ represents the layers, $\alpha= A, B$ represents sublattices, $n$ represents LL index, and $y_c$ is the guiding center. The guiding centers in $T_{2,3}$ ineterlayer hopping terms shift $y_c$ by $\pm\Delta$. Thus, one can write
\begin{equation}
y_c=y_0+(mq+j)\Delta,
\end{equation}
with $j\in{0,1,...,q-1}$, and $\Delta=\sqrt{3}k_\theta \ell^2/2$. 
The moir\'e unit-cell must be commensurate with the magnetic unit-cell such that the associated Hamiltonian becomes diagonalizable for a system of infinite size. The magnetic flux $\phi$ through the unit-cell is given by \cite{bistritzer2011moire}
\begin{equation}
	\phi=\frac{q}{p} \phi_{0}, \quad \phi_{0}=\frac{h c}{e},
\end{equation}
where $p/q \in \mathbb{Q}$ the rational number connecting the size of bare magnetic and moir\'{e} Brillouin zones  when both lattices are commensurate. To be more specific, the end result of the magnetic moir\'{e} Brillouin zone (MMBZ) is limited by
\begin{equation}
	0<k_{x}=\frac{ y_{0}}{\ell^{2}}<\frac{4 \pi p}{q k_{\theta} \ell^{2}}, \quad 0< k_y<\frac{4 \pi}{\sqrt{3}k_{\theta}q}.
\end{equation}
The Fourier transform thus offers a computationally convenient basis
\begin{equation}
	\ket{L,n,\alpha,j}=\frac{1}{\sqrt{N}}\sum_m e^{ik_y (mq+j)\Delta}\ket{L,n,y_0+(mq+j)\Delta}.
	\label{eq:magn_basis}
\end{equation}
We are forced to remove $k_y$ from the basis because the Hamiltonian is diagonal in $k_y$. The intralayer Hamiltonian in terms of the basis in Eq. \ref{eq:magn_basis} is written as
\begin{equation}
	\eqfitpage{h(\theta/2)=\omega_{c} \sum\limits_{L, n, j}\left(e^{-i\theta/2} \sqrt{n+1}\ket{L, n+1, A, j}\bra{L,n, B, j}\right)+\mathrm{H.c.}}
	\label{eq:htheta}
\end{equation}
The interlayer Hamiltonians, on the same basis, are
\small	\begin{align}
	&T_{12}(\mathbf{k})=\sum_{n'n \alpha \beta j}\left[T_{1} F_{n' n}\left(\mathbf{z}_1\right) e^{-i k_x k_{\theta} \ell^2} e^{-4 \pi i \frac{p}{q} j}\ket{2 n' \alpha j}\bra{ 1 n \beta j}\right. \nonumber \\ 
	&+T_{2}F_{n' n}\left(\mathbf{z}_{2}\right) e^{i k_{y} \Delta} e^{\frac{i}{2} k_x k_{\theta} \ell^2} e^{i \pi \frac{p}{q}(2 j-1)}\ket{ 2 n' \alpha, j+1}\bra{ 1 n \beta j} \nonumber \\
	&\left.+T_{3}F_{n' n'}\left(\mathbf{z}_{3}\right) e^{-i k_{y} \Delta} e^{\frac{i}{2} k_x k_{\theta} \ell^2} e^{i \pi \frac{p}{q}(2 j+1)}\ket{ 2 n' \alpha j-1}\bra{ 1 n \beta j}\right],
	\label{eq:ThopmagbasT_12}
\end{align}	
\begin{align}
	T_{23}(\mathbf{k})=&\sum_{n'n \alpha \beta j}\left[T_{1}\ket{3 n' \alpha j}\bra{ 2 n \beta j} 
	+T_{2}\ket{ 3 n' \alpha, j+1}\bra{ 2 n \beta j}\right. \nonumber \\
	&\left.+T_{3}\ket{ 3 n' \alpha j-1}\bra{2 n \beta j}\right],
	\label{eq:ThopmagbasT_23}
\end{align}
with  $ \mathbf{z}_{j}=\frac{q_{j x}+i q_{j y}}{\sqrt{2}}\ell_b$, and 
\begin{equation}
	\begin{aligned}
		F_{n m}(\mathbf{z}) &= \begin{cases}\tilde{F}_{n m}(\mathbf{z}) & n \geq m \\
			\tilde{F}_{n m}^{*}(-\mathbf{z}) & m<n\end{cases} \\
		\tilde{F}_{n m}(\mathbf{z}) &=\sqrt{\frac{m !}{n !}} e^{-\frac{\mathbf{z}^{2}}{2}}\left(-z_{1}+i z_{2}\right)^{n-m} \mathcal{L}_{m}^{n-m}\left(\mathbf{z}^{2}\right)
	\end{aligned}
\end{equation}
where $\mathcal{L}$ is referred to as the associated Laguerre polynomial.
It is crucial to clarify  one subtlety concerning this Hamiltonian's numerical implementation, which was emphasized as a footnote in \cite{bistritzer2011moire}. While the Hamiltonian is simple to execute numerically for the most part, one must be cautious when including states to prevent a false degeneracy at low energies. Consider the case where we are unaware of the interlayer couplings because we chose a model that is relevant near the graphene $K$ point. We must realize that there is only one eigenstate with $K$ point and zero energy per layer. On the other hand, if we choose our fundamental states from
\begin{equation}
	\{L\in\{t,m,b\},\alpha\in\{A,B\},n\in\{0,\dots,n_{\mathrm{max}}\}\},
\end{equation}
and then diagonalize at zero energy, we identify more wrong states. At the $K$ point, we could now relate back to the wavefunctions and analytical equations of the zero energy LL for graphene. At each layer we have  $\ket{n,\pm}=(\pm \ket{n-1},\ket{n})$. The only contributions from sublattice $B$ appear to be $n=0$ in this case.  Since our basis choice does not violate sublattice symmetry, we can assume that the existence of zero energy states with contributions from sublattice $A$ is a numerical artifact. The solutions certainly violate sublattice symmetry, and we must guarantee that this strengthens our numerical method. To do this, a slightly different set of basis states that violate sublattice symmetry is used. Below, we mention such a possibility
\begin{equation}
	\{L\in\{t,m,b\},\alpha\in\{A,B\},n\in\{0,\dots,n_{\mathrm{max}}-\delta_{\alpha,B}\}\}.
\end{equation}
False states are pushed to higher energies as a result of the explicit breakdown of sublattice symmetry in this choice of basis states, but this has no relevance to our situation \cite{bistritzer2011moire}. 
While we emphasize this, when it comes to non-coupling layers, the false low lying levels do not appear. This point is less important in a plot of LLs (because the content does not show degeneracy), but it becomes very helpful in the case of interlayer coupling. Interestingly, LLs are split as interlayer couplings are added, and false low energy bands have a disastrous impact. As a result, it is critical to eliminate the erroneous contributions using the method we just outlined \cite{benlakhouy2022chiral}.
\end{document}